\documentclass{article}
\pdfoutput=1
\usepackage[T1]{fontenc}
\usepackage{multirow}
\usepackage{float}
\usepackage[english]{babel}
\usepackage{amsmath, amssymb, mathtools}
\usepackage{fancyhdr}
\usepackage{scrextend}

\title{\textbf{Clauser-Horne/Eberhard inequality\\
violation by a local model}}
\author{Donald A. Graft\\\textit{donald.graft@cantab.net}}
\date{}
\begin{document}
\lfoot{}
\cfoot{\tiny Copyright \textcopyright { 2016} Donald A. Graft, All rights reserved}
\rfoot{}
\lhead{}
\chead{}
\rhead{}
\maketitle
\thispagestyle{fancy}

\begin{abstract}
Thanks to its immunity to the detection loophole, the Clauser-Horne/\\Eberhard inequality plays an important role in tests of locality and in certification of quantum information protocols based on entanglement. I describe a local model that violates the inequality using a plausible mechanism relying upon a parameter of the apparatus, the source emission rate. The effect is generated through the analysis of time-tagged data using a standard windowed coincidence counting method. Significantly, the detection times here are not functions of the measurement settings, i.e., the `fair coincidences' assumption is satisfied. This finding has implications for the design and interpretation of experiments and for quantum information protocols, as it shows that the coincidence window mechanism cannot be eliminated by a demonstration of independence of the detection times and settings. The paper describes a reliable coincidence counting method and shows that it delivers an accurate count of true coincidences. Recent experimental tests of local realism based on the Clauser-Horne/Eberhard inequality are considered and it is shown that in one case (Christensen et al.) the emission rate is appropriately limited to ensure valid counting, and the data supports locality; in a second case (Giustina et al.) the experiment neglects to appropriately limit the emission rate, and the claimed violation can be accounted for locally. 
\end{abstract}
\bigskip
\textbf{Keywords}: quantum nonlocality, Clauser-Horne/Eberhard inequality, tests of local realism, coincidence window loophole, Einstein-Podolsky-Rosen-Bohm (EPRB) experiments, EPR paradox.

\newpage
\rhead{\thepage}

\section{Introduction}

Inspired by the work of John Bell \cite{Bell00}, inequalities combining the results of several experiments with different measurement settings have been used to distinguish between a fully local universe and one that allows for nonlocal interactions. Significant quantum information protocols, including quantum cryptography, rely on nonlocal correlations (entanglement), and therefore it is important to ensure that the theoretical underpinnings of the protocols are sound. Until recently, all the experimental tests looking for inequality violation have been inconclusive due to their vulnerability to the so-called detection loophole, first described by Pearle \cite{Pearle00}. The detection loophole is triggered when the probability of detection loss depends on the measurement setting. To avoid the detection loophole, Clauser and Horne \cite{ClauserHorne00}, and later Eberhard \cite{Eberhard00}, developed new (logically equivalent) inequalities. When introduced, however, these inequalities required detection efficiencies not reached with available technologies; science had to await the development of detection technologies with adequate efficiency. The recent arrival of adequate technologies, used in conjunction with the CH/Eberhard inequality, has eliminated the detection loophole and enabled proper testing of locality.

Even after the detection loophole is eliminated, concerns remain about the validity of the experimental tests. The ugly truth is that, because the source emissions are not trackable, and because the detections are independently delayed by random amounts, it is not possible to know with certainty whether a given pair of detections in the experimental data is correlated, i.e., whether the two detections result from a single source pair of photons, or whether they represent two single detections from two source pair emissions. A common strategy for resolving this dilemma is to define a pair of detections as coincident when their respective detection times are within a given time interval (the window size) of each other. But this strategy can fail for some window sizes when the detection times are not fully random and instead depend on the measurement settings (this mechanism has come to be known as the `coincidence loophole') \cite{Fine00,Fine01,Scalera,Notarrigo,Pascazio}. To avoid this mechanism, one must assume or arrange that the detection times do not significantly depend on the settings, possibly through proper experimental design and experimental data analyses showing that the times are indeed not modulated by the settings. This assumption of independence of the detection times from the settings is commonly referred to as the `fair coincidences assumption'.

One may feel comforted in believing that one's experiment satisfies the fair coincidences assumption through sound design, and that the experimental data clearly shows no modulation of the detection times by the settings. One may legitimately ask, however, whether adoption of the fair coincidences assumption forecloses all possible local model violations based on coincidence windowing, as is commonly believed, or whether its adoption still leaves open other possibilities for CH/Eberhard violation via mechanisms for which the detection times are not functions of the settings. As a result of researches into this question, I report here, based on theoretical considerations, specification of a simple local model, and a supporting simulation, that the fair coincidences assumption does not foreclose all local violation models based on windowing. The experiments must be re-interpreted and the implications for quantum information protocols must be carefully considered.

To properly appreciate the significance of the results reported here, and to place them in proper context, it is helpful to be acquainted with the history of the locality debate. The review papers \cite{ClauserShimony,Aspect,Genovese00,Genovese01,Brunner,Genovese02} present an important chronology, and in the case of the papers by Genovese, it is intriguing to see how perceptions of the matter have evolved, both generally, and for a single observer.

The plan of the paper is as follows: Section 2 presents a theoretical account of the model demonstrated here, whereby violation of the CH/Eberhard inequality is achieved with purely local operations. Section 3 discusses the methodology for the demonstration of CH/Eberhard inequality violation, including discussion of the CH/Eberhard inequality, its proper normalization, and the importance of the ratio form and positivity measure for assessing the magnitude of violations. Section 4 describes the theoretical operation of the proposed mechanism, and describes and links to the source code of a local computer simulation of the proposed mechanism violating the CH/Eberhard inequality while satisfying the fair coincidences assumption. Section 5 describes the results of the local model simulation. Section 6 discusses the implications of the proposed mechanism for two recent CH/Eberhard-based experiments. Section 7 discusses the significance of the reported findings in a broad context.

\section{Theoretical Basis of the Local Violation}

First define the \textit{CHratio} metric to be used below. Given experiments long enough that the probabilities in the CH inequality \cite{ClauserHorne00} can be represented by ratios of experimental counts, the following \textit{CH} metric and inequality are defined:
\begin{equation*}
\begin{aligned}
CH = C({a_1},{b_1})/N({a_1},{b_1}) + C({a_1},{b_2})/N({a_1},{b_2}) + C({a_2},{b_1})/N({a_2},{b_1})\\
 - C({a_2},{b_2})/N({a_2},{b_2}) - {S_A}({a_1},{b_1})/N({a_1},{b_1}) - {S_B}({a_1},{b_1})/N({a_1},{b_1}) \le 0
\end{aligned}
\end{equation*}

\noindent where \textit{C} denotes the coincidences observed, \textit{S} represents the single counts, and \textit{N} denotes the number of source pair events. Rearranging and dividing produces the \textit{CHratio} metric:
\begin{equation*}
CHratio = \frac{
\splitfrac{C({a_1},{b_1})/N({a_1},{b_1}) + C({a_1},{b_2})/N({a_1},{b_2})}{+ C({a_2},{b_1})/N({a_2},{b_1}) - C({a_2},{b_2})/N({a_2},{b_2})}}
{S_A({a_1},{b_1})/N({a_1},{b_1}) + S_B({a_1},{b_1})/N({a_1},{b_1})
} \le 1
\end{equation*}

For a long enough experiment we can get very close to having the same number of source events in all four arrangements, and so we take $N({a_1},{b_1}) = N({a_1},{b_2}) = N({a_2},{b_1}) = N({a_2},{b_2}) = N$. The counts cancel, simplifying the \textit{CHratio} as follows:
\begin{equation}
CHratio = \frac{{C({a_1},{b_1}) + C({a_1},{b_2}) + C({a_2},{b_1}) - C({a_2},{b_2})}}{{{S_A}({a_1},{b_1}) + {S_B}({a_1},{b_1})}} \le 1
\end{equation}

The single counts are:
\begin{align*}
S_A({a},{b}) = ETs({a})\\
S_B({a},{b}) = ETs({b})
\end{align*}

\noindent where \textit{E} is the emission rate, \textit{T} is the experiment time duration for each measurement arrangement, and \textit{s}() is a function that operates based on the model and the measurement angle (\textit{a} for side A and \textit{b} for side B).

The true coincidence counts are:
\begin{equation*}
{C_{true}}(a,b) = ETc(a,b)
\end{equation*}

\noindent where \textit{c}() is a function of the model and the measurement angles.

Theoretical considerations, as well as dimensional analysis, show that the number of accidental coincidences to be expected in an experiment can be represented as follows:
\begin{equation}
{C_{accidental}} = {E^2}TWd(a,b)
\end{equation}

\noindent where \textit{W} is the coincidence window duration and \textit{d}(a,b) is a function that depends on the model and the measurement settings, and which is independent of \textit{E}, \textit{T}, and \textit{W}. The inclusion of \textit{W} here provides the connection to coincidence window pairing, which is the central concern of this paper.

While this model of accidentals (2) is all that is needed to proceed with the analysis, it is interesting to consider a specific instance of this model. In the course of developing a model for the results of the Giustina et al. experiment, Kofler et al. derive a theoretical expression for the number of accidental coincidences \cite{Kofler}. The expression allows for imperfect detection efficiency and background noise but here I simplify the expression for perfect detection and no noise, as these degrees of freedom are not required to demonstrate the violation and their omission simplifies the presentation. The accidental coincidences are given by:
\begin{equation*}
{C_{accidental}} = {S_A}{S_B}(W/T)(1 - {C_{true}}/{S_A})(1 - {C_{true}}/{S_B})
\end{equation*}

Substituting we have:
\begin{equation*}
{C_{accidental}} = {E^2}TWs(a)s(b)(1 - c(a,b)/s(a))(1 - c(a,b)/s(b))
\end{equation*}

\noindent which has the form of (2). Different models may result in different estimates of accidentals but the key point here is that this analysis does not depend on the precise magnitude of the number of accidentals but rather on the dependence of the number of accidentals on $E^2TW$.

Now form an expression for the \textit{CHratio} metric (1) to be expected from the experiment, with each coincidence count term being the sum of the true coincidences and the accidental coincidences for that arrangement:
\begin{equation*}
\begin{aligned}
CHratio_{exp} = \frac{
\splitfrac
{(ETc({a_1},{b_1}) + {E^2}TWd({a_1},{b_1})) + (ETc({a_1},{b_2}) + {E^2}TWd({a_1},{b_2}))}
{+ (ETc({a_2},{b_1}) + {E^2}TWd({a_2},{b_1})) - (ETc({a_2},{b_2}) + {E^2}TWd({a_2},{b_2}))}}
{
ETs({a_1}) + ETs({b_1})
}
\end{aligned}
\end{equation*}

Rearranging and canceling \textit{ET}:
\begin{equation*}
\begin{aligned}
CHratio_{exp} = \frac{{c({a_1},{b_1}) + c({a_1},{b_2}) + c({a_2},{b_1}) - c({a_2},{b_2})}}{{s({a_1}) + s({b_1})}} &  + \\
\frac{{EWd({a_1},{b_1}) + EWd({a_1},{b_2}) + EWd({a_2},{b_1}) - EWd({a_2},{b_2})}}{{s({a_1}) + s({b_1})}}
\end{aligned}
\end{equation*}

Recognizing that the first expression on the right is the true CH ratio:
\begin{equation*}
\begin{aligned}
CHratio_{exp} = CHratio_{true} + 
EW\bigg[\frac{{d({a_1},{b_1}) + d({a_1},{b_2}) + d({a_2},{b_1}) - d({a_2},{b_2})}}{{s({a_1}) + s({b_1})}}\bigg]
\end{aligned}
\end{equation*}

Simplify:

\begin{equation*}
CHratio_{exp} = CHratio_{true} + EWm({a_1},{a_2},{b_1},{b_2})
\end{equation*}

Consider a classical model that saturates $CHratio_{true}$ at the classical limit of 1. Experimentally, we expect $CHratio_{exp}$ to be less than or equal to 1 for a classical system. Then we have:
\begin{equation}
CHratio_{exp } = 1 + EWm({a_1},{a_2},{b_1},{b_2}) \le 1
\end{equation}

For a given positive non-zero \textit{W} and positive non-zero \textit{m}(), it is clear that to avoid violating the inequality with this classical system we must have \textit{E} equal to zero; otherwise, we can choose the emission rate \textit{E} to obtain an arbitrarily large $CHratio_{exp}$. The simulation to be presented here shows that it is possible to find a classical model for which $CHratio_{true}$ is close to 1 and for which \textit{m}() is positive and non-zero. For such a model, the experimental CH ratio violation is proportional to \textit{E}.

\section{Method for Demonstrating the Local Violation}

Anyone seeking to resolve the problem of unknown true coincidences quickly learns that there are many possible ways to do coincidence pairing when analyzing time-tagged experimental data. It is not immediately obvious which pairing method should be used. One wants to ensure the accuracy of the chosen way, but the true coincidence information is lacking and is therefore not available to validate the possible pairing algorithms. While cogitating on these matters, I realized that when a computer simulation of an Einstein-Podolsky-Rosen-Bohm (EPRB) experiment is executed, the source emissions and resulting detections can be tracked, allowing us to know precisely the true coincidences, while time-tagged data can also be generated and analyzed using multiple algorithms and the results can be compared to the results of true coincidence counting.

While implementing this program, I serendipitously discovered the local model reported here. Accordingly, the present results are derived from a computer simulation, which samples either a classical system or the quantum joint distribution, and which displays the resulting running average CH ratio metric and the running average positivity calculated for both the true coincidences and for the coincidences resulting from a standard moving-window analysis applied to the time-tagged data. Section 4 describes the principles of operation of the model. Complete specification of the simulation is available through perusal of the simulation source code, available online \cite{GraftCode}.

The terms of the CH inequality are probabilities, while the terms of the Eberhard inequality are counts. To obtain probabilities for the CH inequality, the experimental counts must each be divided by the number of emissions (trials) for the respective experiment. But these counts are unknown. This problem can be overcome in practice by enforcing that the emission rate is constant over all the experiments, so that each experiment has the same number of emissions (within acceptable error). Then the CH inequality can be expressed in a ratio form as in Section 2, canceling the number of emissions per trial from the metric. In this ratio form, the CH inequality includes counts and not probabilities, and it is easily seen that the CH and Eberhard inequalities are logically equivalent, justifying our references to the ``CH/Eberhard inequality''. The present analysis uses the ratio form of the CH inequality developed herein as \textit{CHratio} (equation 1). The positivity is also reported.

In practice, the assumption that the four experiments all have the same number of emissions cannot be enforced because the number of emissions for the experiments is stochastic. However, if the experiments are long enough, the stochastic variations become negligible. In the reported simulation, all experiments are run with the same number of emissions to preclude this mechanism.

The denominator of the CH ratio metric is the sum of two singles counts from one of the four experimental arrangements. Singles can be straightforwardly calculated from the time-tagged data with no ambiguity. It is the numerator that needs attention. The numerator is a sum of coincidence counts corresponding to the four experimental arrangements. Three of these counts are weighted positively and one is weighted negatively. The structure of this inequality can therefore be represented as $[c_{a1b1} + c_{a1b2} + c_{a2b1} - c_{a2b2}]$, where the subscripts denote the four experimental arrangements. Validity of the metric requires that the terms of the structure accurately estimate the true coincidence counts. The available pairing methods may inadvertently over-count or under-count the true coincidence counts. If the experimental data analysis over-counts coincidences in all the experiments, then the inequality structure ensures that for a classically saturated system, the inequality can be violated as a consequence of the over-counting. Under-counting, however, would bias the inequality against a violation.

By searching the angle space with a plausible local model that (nearly) saturates the CH inequality, angle sets can be found that violate the inequality using standard coincidence windowing. One suitable angle set, specified in the simulation source code, is selected for use in this report.

\section{Operation of the Local Model Violating CH}

The simulation implements a photon source controlled by a parameter, the source emission rate, which determines the inter-emission times via a Poisson distribution. Detection jitter at the two sides of the experiment is implemented with independent Poisson detection times following the emission time. The detection times are thus independent of the measurement settings. The coincidence window size is set larger than the detection jitter per standard practice to avoid losing true coincidences.

Each source emission is generated with a pair of orthogonal values randomly oriented on 0 to $2\pi$. These paired emission events are distributed one value to each side. The detection apparatus at each side is deterministic and operates as follows:
\\
\\
{\small \fontfamily{ccr}\selectfont \indent for all emission events\\
\indent \indent if cos(a-theta) > 0\\
\indent \indent \indent register a detection event with its detection time}\\

\noindent where a is the measurement angle and theta is the orientation value of the delivered source event.

Maintaining plausible fixed detection jitter values, I investigated the effects of varying the source emission rate and the coincidence window size and found an angle set that violates CH/Eberhard, but only for high source emission rates. There are many such solutions.

The model's windowed coincidence counting algorithm implements a standard moving-window counting method, with a variable coincidence window size. The results show that this counting method is an accurate estimator of the true coincidence counts, but only when the source emission rate is low enough relative to the detection jitter.

The model implements 100 percent detection efficiency without background detections. These physical degrees of freedom are not appealed to for the model's operation and so they are omitted to simplify the model implementation, and to clarify its presentation. The simulation source code can be consulted for a complete specification of the model.

The model uses a coincidence window mechanism to count coincidences; however, the mechanism is fundamentally different from the previously described mechanisms. According to my literature review, Fine deserves the credit for first describing the local coincidence window loophole mechanism \cite{Fine00, Fine01}, which he referred to as a `synchronization' model. Scalera \cite{Scalera} and Notariggo \cite{Notarrigo} later described similar coincidence window models. Fine's mathematical analysis was augmented by Pascazio \cite{Pascazio}. All of the previously described models, including more recent examples, do not satisfy the fair coincidences assumption. Here I show that violation of the fair coincidences assumption is not required to violate the CH/Eberhard inequality using a coincidence window mechanism, and that the fair coincidences assumption is not enough to exclude coincidence window mechanisms that violate CH/Eberhard.

\section{Results of the Local Model Simulation}

Data is gathered from multiple simulation runs with the following conditions: Each individual experiment contains 10,000 emissions and the metrics are assessed over 5000 runs of the four experiments. The Poisson rate factor for the source emission process is tested at three values. The Poisson rate factor for the detection delays is fixed. For the values refer to the simulation source code. The coincidence window size is fixed at a plausible value larger than the detection jitter, so that true coincidences are not lost. The measurement angle set used is specified in the simulation source code. A pseudorandom number generator is used to implement the model's stochasticity. The results are robust to the use of different generators and seeds.

Table 1 shows the CH metrics resulting from both true and windowed coincidence counting with a classical model run with three different source emission rates. The metrics based on true counting are unaffected by differing source emission rates, as one would expect (the true counts are not affected by the emission rate when the total number of emissions is fixed, as here; if the duration of the experiment is fixed then the true counts would increase with the emission rate). The metrics based on windowed counting, however, show clear violations of the CH/Eberhard inequality when the source emission rate is high or very high (a violation is indicated when the CH ratio exceeds 1.0 or the positivity exceeds 0.5). The extent of violation is proportional to the source emission rate; as the emission rate is increased, the violation increases, confirming the theoretical prediction of equation (3).
\\
\\
\centerline {Table 1. CH ratio and positivity for low, high,}
\centerline {and very high source emission rates with a classical model.}
\begin{table}[H]
\centering

\begin{tabular}{|c|c|c|c|c|}
\hline
\multirow{2}{*}{Metric} & \multirow{2}{*}{True counts} & \multicolumn{3}{c|}{Windowed counts}                                                                                            \\ \cline{3-5} 
                        &                              & \multicolumn{1}{l|}{Low} & \multicolumn{1}{l|}{High} & \multicolumn{1}{l|}{Very high} \\ \hline
CH ratio                & 0.999862                     & 0.999706                               & 1.003954                                & 1.029627                                     \\ \hline
Positivity              & 0.491800                     & 0.486400                               & 0.639600                                & 0.996600                                     \\ \hline
\end{tabular}
\end{table}

The results demonstrate the reported coincidence window effect that operates in the absence of dependence between the detection times and the measurement settings, establishing the central claim of this paper. The results also suggest that windowed counting can reliably estimate the true coincidences when the source emission rate is low enough. It is instructive to set a breakpoint in the simulation in order to stop it at an appropriate point to observe the actual coincidence counts. For one run of the four experiments, the following true counts were obtained: [4871, 4021, 4381, 3356]. At a high emission rate the corresponding windowed counts were: [5018, 4839, 4916, 4916]. The coincidences are significantly over-counted for all four experiments, and the structure of the inequality ensures that a violation is produced. In contrast, at a low emission rate the windowed counts were: [4871, 4027, 4383, 3364]. These are much better estimates of the true counts. As the emission rate is lowered, the windowed counts get progressively closer to the true counts. The dependence of the number of over-counted coincidences (accidentals) on the emission rate is derived in Section 2.

To further validate the use of windowed counting with a low source emission rate, the metrics were gathered for a quantum model, and are shown in Table 2. The windowed counting correctly witnesses the expected violation with a magnitude close to that of true counting. Therefore, windowed counting with a low source emission rate is fully adequate to distinguish quantum from classical behavior (assuming also, of course, that the detection efficiencies are high enough and the noise low enough).
\\
\\
\centerline {Table 2. CH ratio and positivity for a low}
\centerline { emission rate with a quantum model.}
\begin{table}[H]
\centering
\begin{tabular}{|c|c|c|}
\hline
Metric     & True counts & Windowed counts \\ \hline
CH ratio   & 1.072339    & 1.070617        \\ \hline
Positivity & 1.000000    & 1.000000        \\ \hline
\end{tabular}
\end{table}

\section{Application to the EPRB Experiments}

All of the EPRB experiments (testing the Bell/CHSH inequality) conducted prior to the availability of high detection efficiencies must be regarded as anach-
ronisms due to their vulnerability to the detection loophole. Plausible variable detection models have been demonstrated for all of them. For example, the Weihs et al. experiment \cite{Weihs} can be accounted for by incorrect calibration of the detector thresholds \cite{GraftWeihs}. One also wonders how an experiment with a claimed 5 percent detection efficiency can seek to establish an inequality violation when the theoretically required detection efficiency is 67 percent \cite{Eberhard00}. None of these low-efficiency experiments qualify as evidence for quantum nonlocality, and they should not be cited in support of it. Only the high-efficiency experiments testing the CH/Eberhard inequality qualify.

Considering the high-efficiency experiments, it is apparent that they are vulnerable to the reported mechanism and this underlines the need for proper control of the source emission rate in the experiments together with an accurate coincidence counting method, such as the coincidence window algorithm used here. Two recent high-efficiency experiments are of great interest: the Christensen et al. experiment \cite{Christensen} and the Giustina et al. experiment \cite{Giustina}.

Christensen et al. claim a small violation of CH. However, their data analysis illegitimately post-selects the data and this post-selection has been shown to account for the reported violation \cite{GraftChristensen}. We are thus motivated to perform an alternate data analysis that eliminates the post-selection. A natural choice is the moving-window pairing method, which while eliminating post-selection, also allows us to assess the role of accidental coincidences in the Christensen et al. experiment. Therefore, I developed an analysis using a variable coincidence window size. First, the detection events were corrected for time-of-flight. Next, detection events outside a Pockels cell opening were removed (they are noise). Then a list of remaining events was created and sorted for each experiment. These lists were analyzed for coincidences using the conventional coincidence counting algorithm with a variable window size.

Based on theoretical models of coincidence counting, it is known that the derivative $C^{\prime} = dC/dW$ of the total number of coincidences with respect to the window size in an experiment is a function of the square of the source pair emission rate $E$. In an experiment where the window size is larger than the detection jitter (to avoid losing true coincidences), if no accidentals are present then $C^{\prime}$ will be 0, because after all the true coincidences are tallied further increases in the window size do not tally any accidentals.

Figure 1 shows the results of the alternative analysis. Shown are the coincidence counts and raw CH metric (this is the CH metric including non-normalized counts, not the ratio metric; it too is valid under the assumption that all the experiments have the same number of emissions) for a windowed analysis of the Christensen et al. experimental data. It can be seen that at a window size greater than about 1000 nanoseconds, all the true coincidences have been tallied and the slopes become very close to 0. This shows that accidentals (over-counted coincidences) are negligible in the Christensen et al. experiment. The CH metric never exceeds 0. While some may argue that it
goes too far to view this as confirming local realism, this failure to confirm quantum nonlocality certainly leaves local realism intact as a viable characterization of nature.
\\
\\
\centerline {Figure 1. Coincidence counts and CH metric versus coincidence}
\centerline {window size in the Christensen et al. experiment.}
\begin{center}\includegraphics[scale=0.4]{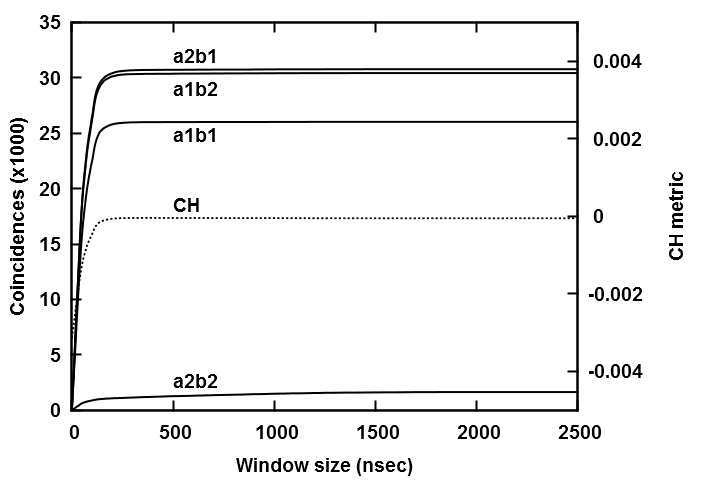}\end{center}
\medskip
The negligible miscounting in the Christensen et al. experiment implies that the experiment was conducted with a suitably low source emission rate, and that the mechanism of this paper is thereby excluded. The Christensen et al. experiment is the first to successfully exclude both the detection and coincidence loopholes.

Figure 2 shows the corresponding analysis for the Giustina et al. experiment. The Eberhard J value is equivalent to the CH metric with all signs reversed, so a value of J below 0 denotes a violation. The accidental coincidences are not negligible, and are especially large in the a2b2 experiment. The non-zero slopes in Figure 2 show that significant miscounting is occurring, indicating that the experiment was conducted with an overly high source emission rate. The mechanism of this paper is thus not excluded and it can account in a local way for the observed violation.
\\
\\
\centerline {Figure 2. Coincidence counts and Eberhard metric versus coincidence}
\centerline {window size in the Giustina et al. experiment.}
\begin{center}\includegraphics[scale=0.4]{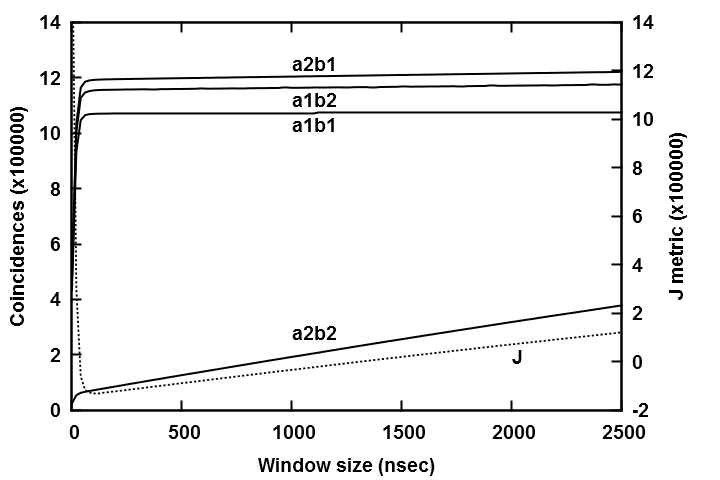}\end{center}

\section{Discussion} 

This result naturally raises doubts about the extent of the application domain of the `no-go' theorems that are cited to underpin quantum information protocols. A reassessment of their significance is called for and some lessons should be learned.

One important lesson to learn from the simulation results is that the data analysis must be sound, such that coincidences are not significantly over-counted. Trying to correct for bad counting using ad hoc models for accidentals, etc., is dangerous and unnecessary. A simple, sound, and robust analytical method that can be recommended is the moving-window analysis described here, together with a low experimental source emission rate. This important caveat on source emission rates was long ago described by Clauser and Horne \cite{ClauserHorne00}:
\\
\begin{addmargin}[2em]{2em}
`Thus, we tacitly require the experimental arrangement to be such that this condition obtains (suitable source strength, time separation of pairs, etc.). If a sufficiently weak source is used, the ratio of ``chance'' coincident counts to ``true'' coincident counts can be made arbitrarily small, and the corresponding dead time can also be minimized.'
\\
\end{addmargin}

Clauser and Horne required that the average time between source pair emission events be much greater than the coincidence window size. Clauser and Horne do not explicitly state the consequences of using an overly high emission rate beyond stating that doing so can lead to a large number of accidentals, i.e., they do not state the effect of these accidentals on the CH metric. Subsequently, researchers have incorrectly assumed that accidentals (over-counted coincidences) must reduce the CH metric. For example Zukowski writes as follows \cite{Zukowski}:
\\
\begin{addmargin}[2em]{2em}
``Only counts related with true entangled pairs may lead to violation of a Bell inequality. If more than one pair is emitted from a single pulse, this lowers the visibility (interferometric contrast) of the observed two detector interference. Thus, if the observed two-detector interference is good enough to violate a Bell inequality, then we have an indirect indication that the number of double-pair emissions was low.''
\\
\end{addmargin}

Consequently, the requirement for a low emission rate has been disregarded and experiments are typically performed without controlling the source emission rate. A notable exception is the Christensen et al. experiment \cite{Christensen}, which, when properly interpreted, offers experimental evidence disconfirming quantum nonlocality. Here I showed that an excessive emission rate can indeed increase the CH metric (due simply to the structure of the inequality coupled with over-counting of coincidences), and I presented an alternative way to correctly and practically calibrate the emission rate: reduce it until the coincidences versus window size curves become flat.

Another important lesson is that coincidence models that satisfy the fair coincidences assumption may nevertheless violate the CH/Eberhard inequality. This has not been demonstrated until now and all previous coincidence models rely upon a dependence of the detection times on the measurement settings, i.e., they rely upon violation of the fair coincidences assumption. It is not enough to claim that in an experiment the detection times do not depend on the settings.

I have previously argued that CH/Eberhard inequality violation entails nonlocal transfer of settings information between the sides \cite{GraftChristensen}. The violation achieved with the model of this paper does not challenge that notion because the entailment depends on accurate estimation of the true coincidences. If the estimates are accurate and violation is achieved, then some covert information sharing is implied. The validity of the CH/Eberhard inequality depends on valid coincidence counting in the experimental data analyses.

In the future, I hope to apply the methodology developed here to an analysis and assessment of the different possible counting methods, including full counting \cite{GraftChristensen}, pulsed source counting \cite{Christensen}, Giustina-style counting \cite{Giustina}, recent distance-based measures \cite{Knill}, and others.

\section*{Acknowledgements} 

I thank the Christensen et al. and Giustina et al. teams for making available the raw data and analyses from their experiments. I thank Arthur Fine for ongoing useful discussions in several areas of quantum physics and philosophy, and for his encouragement of my pursuit of local realistic understanding. I thank Marco Genovese for wise and generous guidance.



\begin{thebibliography}{99}

\bibitem{Bell00} J. S. Bell, {\it Speakable and Unspeakable in Quantum Mechanics},
2nd ed., Cambridge University Press, Cambridge (1987).

\bibitem{Pearle00} P. Pearle, ``Hidden-variable example based upon data rejection'', {\it Physical Review D} {\bf 2}, 1418-1425 (1970). 

\bibitem{ClauserHorne00} J. F. Clauser and M. A. Horne, ``Experimental consequences of objective local theories'', {\it Physical Review D} {\bf 10}(2), 526-535 (1974).

\bibitem{Eberhard00} P. H. Eberhard, ``Background level and counter efficiencies required for a loophole-free Einstein-Podolsky-Rosen experiment'', {\it Physical Review A} {\bf 47}(2), R747-R750 (1993).

\bibitem{Fine00} A. Fine, ``Correlations and Physical Reality'', Proceedings of the Biennial Meeting of the Philosophy of Science Association {\bf 1980}(2), 535-562 (1980).

\bibitem{Fine01} A. Fine, ``SOME LOCAL MODELS FOR CORRELATION EXPERIMENTS'', {\it Synthese} {\bf
50}, 279-294 (1982).

\bibitem{Scalera} G. C. Scalera, ``On a Local Hidden-Variable Model with Unusual Properties'', {\it Lett. Nuovo Cimento} {\bf 38}(1), 16-18 (1983).

\bibitem{Notarrigo} S. Notarrigo, ``A Newtonian Separable Model which Violates Bell's Inequality'', {\it Il Nuovo Cimento B} {\bf 83}(2), 173-187 (1984).

\bibitem{Pascazio}S. Pascazio, ``TIME AND BELL-TYPE INEQUALITIES'', {\it Phys. Lett. A} {\bf 118}(2), 47-53 (1986).

\bibitem{ClauserShimony} J. F. Clauser and A. Shimony, ``Bell's theorem: experimental tests and implications'', {\it Rep. Prog. Phys.}, {\bf 41}, 1881-1927 (1978). 

\bibitem{Aspect} A. Aspect, ``BELL'S THEOREM: THE NAIVE VIEW OF AN EXPERIMENTALIST'', in ``Quantum [Un]speakables -- From Bell to Quantum information'', edited by R.A. Bertlmann and A. Zeilinger, Springer (2002). 

\bibitem{Genovese00} M. Genovese, ``Research on hidden variable theories: A review of recent progresses'', {\it Physics Reports} {\bf 413}(6), 319-396 (2005).

\bibitem{Genovese01} M. Genovese, ``Experimental tests of Bell inequalities'', {\it International Journal of Quantum Foundations},  \small\fontfamily{ccr}\selectfont http://www.ijqf.org \normalfont \normalsize (2014).

\bibitem{Brunner} N. Brunner, D. Cavalcanti, S. Pironio, V. Scarani, and S. Wehner, ``Bell nonlocality'', {\it Rev. Modern Phys.} {\bf 86}(419), (2014); Erratum {\it Rev. Modern Phys.} {\bf 86}(839), (2014).

\bibitem{Genovese02} M. Genovese, ``50 Years of Experimental Bell Inequalities: A Short Introduction'', {\it Journal of Advanced Physics} {\bf 4}(3), 233-235 (2015).

\bibitem{Kofler} J. Kofler, S. Ramelow, M. Giustina, and A. Zeilinger, ``On `Bell violation using entangled photons without the fair-sampling assumption'\thinspace'', arXiv:1307.6475 (2013).

\bibitem{GraftCode} \begin{flushleft}
D. A. Graft, the simulation source code is available here: \small \fontfamily{ccr}\selectfont http://rationalqm.us/papers/Coincidence window model/CH Violation.cpp\end{flushleft} \normalfont \normalsize


\bibitem{Weihs} G. Weihs, T. Jennewein, C. Simon, H. Weinfurter, and A. Zeilinger, ``Violation of Bell's inequality under strict Einstein locality conditions'', {\it Phys. Rev. Lett.} {\bf 81}, 5039-5043 (1998).

\bibitem{GraftWeihs} D. A. Graft, ``A local realist account of Weihs, Jennewein, Simon, Weinfurther, and Zeilinger EPRB experiment'', {\it Physics Essays} {\bf 26}(2), 214-222 (2013).

\bibitem{Christensen} B. G.	Christensen, K. T. McCusker, J. B. Altepeter, B. Calkins, C. C. W. Lim, N. Gisin, and P. G. Kwiat, ``Detection-Loophole-Free Test of Quantum Nonlocality, and Applications'', {\it Phys. Rev. Lett.} {\bf 111}, 130406 (2013).

\bibitem{Giustina} M. Giustina, A. Mech, S. Ramelow, B. Wittmann, J. Kofler, A. Lita, B. Calkins, T. Gerrits, Sae Woo Nam, R. Ursin, and A. Zeilinger, ``Bell violation with entangled photons, free of the fair-sampling assumption'', {\it Nature} {\bf 497}, 227 (2013).  

\bibitem{GraftChristensen} D. A. Graft, ``Analysis of the Christensen et al. Test of Local Realism'', {\it J. Adv. Phys.} {\bf 4}(3), 284-300 (2015).

\bibitem{Zukowski} M. Zukowski, private communication (2015).

\bibitem{Knill} E. Knill, S. Glancy, Sae Woo Nam, K. Coakley, and Yanbao Zhang, ``Bell inequalities for continuously emitting sources'', {\it Phys. Rev. A} {\bf 91}(3), 032105 (2015).

\end{thebibliography}
\end{document}